\documentclass[aps, prb, showpacs, twocolumn, 
amssymb,superscriptaddress]{revtex4}
\usepackage[dvipdfmx]{graphicx}
\usepackage{bm, amsmath, amsfonts}
\usepackage{color}

\def\ra{\rangle}
\def\la{\langle}

\def\dag{\dagger}
\def\fc{\mathfrak{c}}
\def\fs{\mathfrak{s}}
\newcommand{\ii}{\text{i}}
\def\Oe{O_{\scalebox{0.95}{e}}}
\def\Oo{O_{\scalebox{0.95}{o}}}
\def\Psp{\Psi^{(+)}_0}
\def\Psm{\Psi^{(-)}_0}
\def\Pspm{\Psi^{(\pm)}_0}
\def\Psmp{\Psi^{(\mp)}_0}
\def\Pse{\Psi^{({\rm even})}_0}
\def\Pso{\Psi^{({\rm odd})}_0}

\begin{document}


\title{
Exact ground states and topological order in interacting Kitaev/Majorana chains
}

\author{Hosho Katsura}
\affiliation{
Department of Physics, Graduate School of Science, The University of Tokyo, 
Hongo, Tokyo 113-0033
}

\author{Dirk Schuricht}
\affiliation{
Institute for Theoretical Physics, Center for Extreme Matter and Emergent Phenomena,
Utrecht University, Leuvenlaan 4, 3584 CE Utrecht, The Netherlands
}

\author{Masahiro Takahashi}
\affiliation{
Department of Physics, Gakushuin University, Tokyo 171-8588, Japan
}

\date{\today}
\begin{abstract}
We study a system of interacting spinless fermions in one dimension which, in the absence of interactions, reduces to the Kitaev chain [A. Yu Kitaev, Phys.-Usp. \textbf{44}, 131 (2001)]. In the non-interacting case, a signal of topological order appears as zero-energy modes localized near the edges. We show that the exact ground states can be obtained analytically even in the presence of nearest-neighbor repulsive interactions when the on-site (chemical) potential is tuned to a particular function of the other parameters. 
As with the non-interacting case, the obtained ground states are two-fold degenerate and differ in fermionic parity. We prove the uniqueness of the obtained ground states and show that they can be continuously deformed to the ground states of the non-interacting Kitaev chain without gap closing. We also demonstrate explicitly that there exists a set of operators each of which maps one of the ground states to the other with opposite fermionic parity. These operators can be thought of as an interacting generalization of Majorana edge zero modes.
\end{abstract}


\maketitle

\section{Introduction}
\label{sec:intro}

In recent years Majorana zero modes have been in the focus of research in condensed matter physics.\cite{Wilczek09,Alicea12,LeijnseFlensberg12,Beenakker13} 
Experimental signatures of Majorana zero modes have been observed in the tunnelling conductance in  hybrid superconductor-semiconductor nanowire systems.\cite{Mourik-12,Rokhinson-12,Das-12,Deng-12} 
However, an unambiguous identification of Majorana particles remains open.\cite{Churchill-13} 
Very recently an important step~\cite{Nadj-Perge-14} towards such an identification was taken by scanning tunnelling microscopy experiments on ferromagnetic atomic chains on a superconductor, which allowed to obtain spatially resolved signatures showing localization of the zero modes at the edges as predicted by theory. 

The prototypical toy model possessing Majorana zero modes is the Kitaev chain with open boundaries,\cite{Kitaev01} 
a one-dimensional tight-binding model for spinless fermions in the presence of p-wave superconducting pairing. 
This model possesses a topological phase with two-fold degenerate ground states which cannot be distinguished by any local order parameter. 
There exist zero-energy modes that commute with the Hamiltonian and anticommute with the fermionic parity; these are the Majorana zero modes. 
They are exponentially localized near the boundaries. 
As is well-known (see e.g. Ref.~\onlinecite{Fendley12}) the Kitaev chain can be mapped to the one-dimensional transverse-field Ising chain via a non-local Jordan--Wigner transformation. In the resulting spin variables, the topological phase reduces to the ordered phase showing spontaneous magnetization while the fermionic parity maps to a string of spin operators. 

The recent surge of interest in condensed-matter realizations of Majorana fermions 
includes the investigation of the effects of disorder~\cite{Brouwer-11,Lobos-12,DeGottardi-14,Altland-14,Crepin-14} or dimerization~\cite{Wakatsuki-14} on the topological phase, the study of nanostructures possessing Majorana zero modes,\cite{LeijnseFlensberg11,Walter-11,Huetzen-12,Leijnse14,Lopez-14} and strongly correlated systems showing Kondo physics.\cite{BeriCooper12,Tsvelik13,Altland-14b,Cheng-14,Galpin-14,Eriksson-14} Furthermore, generalizations to higher symmetries including supersymmetry~\cite{Tsvelik12,Grover-14,UASH14,Rahmani-15-1,Rahmani-15-2} and parafermion modes were analyzed.\cite{Fendley12,Clarke-14,KlinovajaLoss14,Mong-14,Jermyn-14, Aris-15} 

Here we focus on interaction effects. Motivated by the proposal~\cite{Lutchyn-10,Oreg-10} to realize systems hosting Majorana zero modes in semiconducting nanowires and the subsequent experimental works,\cite{Mourik-12,Rokhinson-12,Das-12,Deng-12,Churchill-13} most previous studies of interaction effects have focused on nanowire setups. This included the original spinful nanowire system with spin-orbit coupling, a Zeeman field and proximity to a superconductor,\cite{Gangadharaiah-11,Stoudenmire-11,Manolescu-14} similar multi-band nanowires,\cite{LutchynFisher11} and effectively spinless systems~\cite{Lobos-12,Crepin-14} including helical wires~\cite{Sela-11} and two-chain ladders.\cite{ChengTu11} Furthermore, interactions directly in the Kitaev chain were studied,\cite{Gangadharaiah-11,HS12,Thomale-13,Rahmani-15-1,Milsted-15,Rahmani-15-2, KueiSun-15} which were also shown to be implementable in an array of Josephson junctions,\cite{HS12} a realization allowing to reach strong interaction strengths as well as good control of the parameters. From all these works a physical picture emerged revealing two main effects of interactions: On the one hand, interactions suppress the bulk gap and thus decrease the stability of the topological phase, while on the other hand, repulsive interactions broaden the chemical-potential window over which Majorana zero modes exist. Which of these two effects dominates or is more relevant depends on the specific realization at hand. General aspects of interaction effects on topological phases were also analyzed in Refs.~\onlinecite{FidkowskiKitaev10,FidkowskiKitaev11, Turner11, Gurarie11,GoldsteinChamon12,Kells14,Chiu-15}.

In this article, we further investigate the Kitaev chain in the presence of repulsive interactions between the spinless fermions. Employing the Jordan--Wigner transformation, this model can be mapped to an XYZ Heisenberg chain, allowing to obtain a wealth of results on the phase diagram.\cite{Sela-11, HS12} Here, however, we are not interested in the phase diagram but directly in the two-fold degenerate ground states in the topological phase. Using classic results in the spin chain literature,\cite{PeschelEmery81,Kurmann-82,MuellerShrock85} we find that the exact ground states of the model can be obtained when the chemical potential is tuned to a particular function of the other parameters: the hopping amplitude, the p-wave pairing gap, and the strength of the repulsive interaction.  In this special case, the ground states can be written in a simple product form. 
The ground-state degeneracy is a necessary but not sufficient condition for topological order and the existence of Majorana zero modes. In order to prove 
the existence of topological order, we show that there exists a smooth path that interpolates between the solvable Hamiltonian and the non-interacting Kitaev chain in the topological phase, along which the ground states remain unchanged. 
We also prove rigorously that the spectral gap above the ground states does not vanish along the entire path. 
Then, thanks to the results of Fidkowski and Kitaev,\cite{FidkowskiKitaev11} it follows that the topological order of the non-interacting Kitaev chain persists along the path. 
To provide further evidence for the topological order, we prove the existence of a set of operators each of which maps one of the ground states to the other with the opposite fermionic parity. The operators obtained (i) are hermitian, (ii) anticommute with the fermionic parity, and (iii) are localized near the edges. Therefore, they can be regarded as an interacting generalization of the Majorana zero modes. In fact, in the absence of the interactions, they exactly commute with the Hamiltonian and reduce to the standard Majorana zero modes in the original Kitaev chain. 

The paper is organized as follows. 
In Sec. \ref{sec:model}, we give a precise definition of the model and list the symmetries of its Hamiltonian. In Sec. \ref{sec:gs}, we first introduce the notion of frustration-free Hamiltonians. Then we show how one can find the ground states of a chain of arbitrary length from the exact results of the two-site problem. The expressions for the ground states are also given explicitly. In Sec. \ref{sec:topo order}, we discuss the topological order of the present system. We present a theorem about the continuous deformation between the interacting and non-interacting Kitaev chains that share the same ground states. 
We show how the theorem follows from a lemma about the auxiliary free-fermion problem. 
We also present the proof of the lemma and derive explicit expressions for the Majorana zero modes in the auxiliary free-fermion problem. 
Concluding remarks are presented in Sec. \ref{sec:conc}. 
In App.~\ref{sec:Majo Ham}, we summarize the relation between Majorana and complex fermions. The explicit expression for the Hamiltonian in terms of Majorana fermions is also given. In App.~\ref{sec:spin Ham}, we show the mapping between the fermionic Hamiltonian and the XYZ chain in a magnetic field. We also show how the frustration-free condition and the ground states obtained translate into the spin language. 
In App. C, we present a derivation of the inequality used in Sec. III. A few examples of correlation functions in the ground states are also provided. In App. D, we extend the theorem to the case where the couplings are spatially inhomogeneous.
In App.~\ref{sec:evC}, we present a detailed exposition of the eigenvalue problem related to the auxiliary free-fermion problem.

\section{Model}
\label{sec:model}

We consider a system of spinless fermions on a chain of length $L$ with open boundaries. 
For each site $j=1,2, \cdots, L$, we denote by $c^\dagger_j$ and $c_j$ 
the creation and the annihilation operators respectively. 
As usual, the number operators are defined by $n_j:=c^\dagger_j c_j$. 

\subsection{Hamiltonian}

We consider the Hamiltonian of interacting spinless fermions described by
\begin{align}
H &= \sum^{L-1}_{j=1}
      \left[ -t (c^\dagger_j c_{j+1}+c^\dagger_{j+1}c_j)
      +\Delta (c_j c_{j+1}+c^\dagger_{j+1}c^\dagger_j) \right] \nonumber\\
   &-\frac{1}{2}\sum^L_{j=1} \mu_j (2n_j-1)
      +U \sum^{L-1}_{j=1} (2n_j-1) (2n_{j+1}-1),
      \label{eq:Ham1}
\end{align}
where $t$ is the hopping amplitude and $\Delta$ the p-wave paring gap, which is assumed to be real. Without loss of generality, we can assume that $t \ge 0$ because 
the case with $t \le 0$ can be achieved by local unitary transformations: 
$c_j\to-\ii (-1)^jc_j$. 
We can further assume that $\Delta \ge 0$ because the case with $\Delta \le 0$ can be achieved by $c_j \to \ii c_j$ for all $j$. 
Here, $\mu_j$ is the on-site (chemical) potential and $U \ge 0$ is the strength of the nearest-neighbor repulsive interaction. It is convenient, for later purposes, to keep a site-dependent on-site potential $\mu_j$. In the absence of the interaction ($U=0$), the model reduces to the Kitaev chain,\cite{Kitaev01} in which Majorana edge zero modes occur provided that it is in the topological phase. 
Thus one can think of the interacting system as Majorana fermions with a quartic interaction. An explicit expression for the Hamiltonian in terms of Majorana fermions is given in App.~\ref{sec:Majo Ham}. 
One can also express the Hamiltonian in terms of spin-1/2 operators via a Jordan--Wigner transformation. The corresponding model turns out to be the XYZ spin chain in a magnetic field (see App.~\ref{sec:spin Ham} for more details). 

\subsection{Symmetries}
\label{sec:sym}

Let us first consider the symmetries of the Hamiltonian $H$. Because of the presence of the pairing term, $H$ does not conserve the total fermion number $F:= \sum^L_{j=1} n_j$, i.e., $[H,F]\neq 0$. However, the fermion number modulo two is conserved, $[H,(-1)^F]=0$. 
The Hamiltonian respects time-reversal symmetry, i.e., is invariant under complex conjugation.  

In addition, when $\mu_j=0$ for all $j$, the Hamiltonian is invariant under the charge conjugation operation $c_j \to (-1)^j c^\dagger_j$. More precisely, $H$ commutes with the following unitary operator
\begin{equation}
P = \prod^L_{j=1} \left[c_j +(-1)^j c^\dagger_j\right].
\end{equation}
We also note that this particular case is integrable, because it can be mapped to the XYZ spin chain (without magnetic fields) which was solved by Baxter.\cite{Baxter_book} 
For the case with non-vanishing $\mu_j$'s, the Hamiltonian $H$ is not invariant under the above charge conjugation. 
However, we can at least say that $H$ with $\{ \mu_j \}^L_{j=1}$ and that with $\{ -\mu_j \}^L_{j=1}$ have the same spectrum. In the following, for simplicity, we assume that $\mu_j \ge 0$ for all $j$.

\section{Exact ground states}
\label{sec:gs}

In this section we show that the Hamiltonian $H$ is {\it frustration-free} when the $\mu_j$'s are tuned to a particular function of the other parameters $(t, \Delta, U)$, in which case the exact ground states are easy to obtain analytically. 

Consider the case where $\mu_1 = \mu_L = \mu/2$ and $\mu_j=\mu$ ($j=2,3,...,L-1$), i.e., the on-site potentials at the boundary sites are half the bulk ones. 
In this case, the Hamiltonian takes the form
\begin{equation}
H = \sum^{L-1}_{j=1} h_j,
\label{eq:ham1}
\end{equation}
where
\begin{align}
h_j &= -t (c^\dagger_j c_{j+1} + c^\dagger_{j+1} c_j) 
+\Delta (c_j c_{j+1} + c^\dagger_{j+1} c^\dagger_j) \nonumber \\
& - \frac{\mu}{2} (n_j + n_{j+1}-1) +U (2n_j-1) (2n_{j+1}-1).
\end{align}
Since $[ h_j, h_k ] \ne 0$ if $|j-k|=1$, the local Hamiltonians cannot be diagonalized simultaneously. However, there may exist some special values of $\mu$ at which 
the projection onto the ground-state subspace $P_0$ satisfies
$h_j P_0 = \epsilon_0 P_0$ for all $j$, where $\epsilon_0$ is the smallest eigenvalue of $h_j$. In other words, the ground states of $H$ minimize each $h_j$ independently. 
In such cases, the Hamiltonian $H$ is said to be frustration-free. 
Many known exactly solvable models fall into this category. Examples include 
the Affleck-Kennedy-Lieb-Tasaki model,\cite{AKLT} 
the Rokhsar-Kivelson model,\cite{RK_PRL} and the Kitaev toric code model.\cite{Kitaev_toric_code}

To search for the condition under which the Hamiltonian \eqref{eq:ham1} is frustration-free, we shall first consider the two-site problem, i.e., deriving the spectrum and the eigenstates of $h_j$. 
Let us ignore, for the moment, the region outside the bond $(j, j+1)$. 
Then the Hilbert space for the problem is spanned by the four states:
$|\circ \circ\, \rangle := |{\rm vac}\rangle$, 
$|\bullet\circ\,\rangle := c^\dagger_j |{\rm vac}\rangle$,
$|\circ\bullet\,\rangle := c^\dagger_{j+1} |{\rm vac}\rangle$, and 
$|\bullet \bullet\,\rangle := c^\dagger_j c^\dagger_{j+1} |{\rm vac }\rangle$, 
where $|{\rm vac}\rangle$ is the vacuum state such that $c_k |{\rm vac}\rangle=0$ for any $k$.
Since the local Hamiltonian $h_j$ commutes with the fermionic parity $(-1)^F$, we can deal with even and odd sectors separately. In the even sector spanned by $|\circ \circ\, \rangle$ and $|\bullet \bullet\,\rangle$, $h_j$ is expressed as
\begin{equation}
\begin{array}{ccc}
& & |\circ \circ\, \rangle~~~~~|\bullet \bullet\,\rangle \\[2mm]
h^{({\rm even})}_j & = &
\begin{pmatrix}
U+ \mu/2 & -\Delta \\
-\Delta & U-\mu/2
\end{pmatrix}.
\end{array}
\end{equation}
An unnormalized ground state of $h^{({\rm even})}_j$ and the corresponding eigenvalue are
\begin{align}
|\psi^{({\rm even})}_0\rangle &=  
|\circ \circ\, \rangle + \cot \frac{\theta}{2} |\bullet \bullet\,\rangle, \\
\epsilon^{({\rm even})}_0 &= U-\sqrt{\Delta^2+(\mu/2)^2},
\end{align}
respectively, where $\theta = \arctan (2\Delta/\mu)$ which is assumed to be in $[0, \pi]$.  
Note that $|\psi^{({\rm even})}_0\rangle$ is non-degenerate as long as either $\Delta$ or $\mu$ is non-vanishing. 
Similarly, in the odd sector spanned by $|\bullet\circ\,\rangle$ 
and $|\circ\bullet\,\rangle$, $h_j$ is expressed as 
\begin{equation}
\begin{array}{ccc}
& & |\bullet\circ\,\rangle\,|\circ\bullet\,\rangle \\[2mm]
h^{({\rm odd})}_j & = &
\begin{pmatrix}
-U & -t \\
-t & -U
\end{pmatrix}.
\end{array}
\end{equation}
An unnormalized ground state of $h^{({\rm odd})}_j$ and the corresponding eigenvalue are
\begin{align}
|\psi^{({\rm odd})}_0 \ra &= |\bullet\circ\,\rangle + |\circ\bullet\,\rangle, \\
\epsilon^{({\rm odd})}_0 &= -(U+t),
\end{align}
which is non-degenerate except when $t = 0$. 

The absolute ground state of $h_j$ can be obtained by comparing $\epsilon^{({\rm even})}_0$ and $\epsilon^{({\rm odd})}_0$, which become degenerate if 
\begin{equation}
\mu = \mu^* := 4 \sqrt{ U^2+tU+\frac{t^2-\Delta^2}{4} }.
\label{eq:FFcond}
\end{equation}
Note that though $\mu=-\mu^*$ is also allowed, we have chosen the positive $\mu$ for simplicity (see the discussion in Sec. \ref{sec:sym}). 
At $\mu=\mu^*$, any linear combination of $|\psi^{({\rm even})}_0 \ra$ and $|\psi^{({\rm odd})}_0 \ra$ is also a ground state of $h_j$. If we take the following particular combinations, we see that they are {\it disentangled}
\begin{align}
|\psi^{(\pm)}_0 \ra =& |\psi^{({\rm even})}\rangle 
\pm \sqrt{\cot \frac{\theta^*}{2}}\, |\psi^{({\rm odd})}\rangle \nonumber \\
= & (1\pm \alpha c^\dag_j) (1 \pm \alpha c^\dag_{j+1}) |{\rm vac}\rangle \nonumber \\ 
= & \exp (\pm {\alpha c^\dag_j}) \, \exp (\pm{\alpha c^\dag_{j+1}}) |{\rm vac}\rangle,
\end{align}
where $\theta^*=\arctan (2\Delta/\mu^*)$ and $\alpha = \sqrt{\cot (\theta^*/2)}$. We recall that $\theta \in [0, \pi]$ has been assumed. 
In the entire Hilbert space, the ground states of $h_j$ at $\mu=\mu^*$ are highly degenerate, because any state of the form
\begin{equation}
f (c^\dag_1, ..., c^\dag_{j-1})\, 
e^{\scalebox{0.95}{$\pm \alpha c^\dag_j$}} \, 
e^{\scalebox{0.95}{$\pm \alpha c^\dag_{j+1}$}}\,
g(c^\dag_{j+2}, ..., c^\dag_{L}) |{\rm vac}\ra
\end{equation}
is a ground state of $h_j$, where $f$ and $g$ are arbitrary polynomials 
in $c^\dag_1, ..., c^\dag_{j-1}$ and $c^\dag_{j+2}, ..., c^\dag_{L}$, respectively. 

In this way, we find that the ground states of $h_j$ at $\mu=\mu^*$ can be factorized into the product of two states. This observation hints at the possibility that we may obtain exact ground states of $H$ with $\mu=\mu^*$ for any $L$. In fact, $H$ is frustration-free at $\mu=\mu^*$. 
To see this, consider the states of the form
\begin{align}
|\Psi^{(\rm \pm)}_0 \ra &= \frac{1}{(1+\alpha^2)^{L/2}}\, 
e^{\scalebox{0.95}{$\pm \alpha c^\dag_1$}} 
e^{\scalebox{0.95}{$\pm \alpha c^\dag_2$}} \cdots
e^{\scalebox{0.95}{$\pm \alpha c^\dag_L$}}  
|{\rm vac}\ra.
\label{eq:gs1}
\end{align}
Here the coefficient has been introduced so that $\la \Psi^{(\pm)}_0|\Psi^{(\pm)}_0\ra=1$.
Since $\exp (\pm \alpha c^\dag_k)$ commutes with $h_j$ unless $k \ne j, j+1$, one can easily verify that $|\Psi^{(\rm \pm)}_0 \ra$ minimizes each $h_j$ independently at $\mu=\mu^*$. Therefore, $H$ is frustration-free at $\mu=\mu^*$ and $|\Psi^{(\rm \pm)}_0 \ra$ are its exact ground states with energy\cite{RefB}
\begin{equation}
E_0 = -(L-1) (U+t).
\end{equation}
A similar factorization of the ground states has been found in spin-$1/2$ chains in a magnetic field.\cite{PeschelEmery81,Kurmann-82,MuellerShrock85} In fact, the condition under which the Hamiltonian is frustration-free is the same as the exactly solvable condition of the XYZ spin chain in a magnetic field (see App.~\ref{sec:spin Ham} for more details). 

It should be noted that $|\Psi^{(\pm)}_0\ra$ are not orthogonal and are not eigenstates of the fermionic parity $(-1)^F$, because each $|\Psi^{(\pm)}_0 \ra$ contains both even and odd numbers of fermions. 
The ground states that are eigenstates of $(-1)^F$ can be obtained by taking appropriate linear combinations of $|\Psi^{(\pm)}_0\ra$. Let us decompose the entire Hilbert space ${\cal H}$ as ${\cal H} = {\cal H}^{({\rm even})} \oplus {\cal H}^{({\rm odd})}$, where ${\cal H}^{(\rm even/odd)}$ consists of the states with even/odd numbers of fermions. Then, noting that $[ F, c^\dag_j] = c^\dag_j$ and hence $(-1)^F \exp (\pm \alpha c^\dag_j) = \exp (\mp \alpha c^\dag_j) (-1)^F$, we find that the following states
\begin{align}
|\Psi^{(\rm even)}_0 \ra &= |\Psi^{(+)}_0 \ra + |\Psi^{(-)}_0 \ra \in {\cal H}^{({\rm even})}, 
\label{eq:egs}\\
|\Psi^{(\rm odd)}_0 \ra &=|\Psi^{(+)}_0 \ra - |\Psi^{(-)}_0 \ra \in {\cal H}^{({\rm odd})},
\label{eq:ogs}
\end{align}
are orthogonal to each other. 
As we will prove in the next section, there is no other ground state
of $H$ with $\mu=\mu^*$ except for these two states. 
Furthermore, for large $L$, they cannot be distinguished by any local measurement. 
To illustrate this, let us consider the expectation values of local operators in $|\Psi^{(\rm even/odd)}_0 \ra$. For notational convenience, we write
\begin{equation}
\langle  \cdots \rangle_{\rm par} :=  
\frac{\langle \Psi^{({\rm par})}_0 | \cdots | \Psi^{({\rm par})}_0 \rangle}{\langle \Psi^{({\rm par})}_0 | \Psi^{({\rm par})}_0 \rangle},
\end{equation}
where ${\rm par}$ is either ${\rm even}$ or ${\rm odd}$. 
Let $\Oe/\Oo$ be a local operator consisting of even/odd number of creation and annihilation operators. We assume that $\Oe/\Oo$ is supported on a set of lattice sites $j_1 < j_2 < \cdots <j_k$ with $j_k-j_1 = \ell-1$. 
Noting that $\{ \Oo, (-1)^F \} =0$ and using the relations $(-1)^F\, |\Pse\ra = |\Pse\ra$ and $(-1)^F\, |\Pso\ra = -|\Pso\ra$, one finds
\begin{equation}
\la \Oo \ra_{\rm even} = \la \Oo \ra_{\rm odd} =0,
\label{eq:expec1}
\end{equation}
for any $\Oo$. On the other hand, for $\Oe$, one finds that the difference between $\la \Oe \ra_{\rm even}$ and $\la \Oe \ra_{\rm odd}$ is bounded from above as follows:
\begin{equation}
\big| \la \Oe \ra_{\rm even} - \la \Oe \ra_{\rm odd} \big| \,\le\,  {\cal C}\, \| \Oe \|\, e^{\scalebox{0.90}{$-L/\xi$}},
\label{eq:expec2}
\end{equation}
where $\| \Oe \|$ denotes the operator norm of $\Oe$,
\begin{equation}
\xi = \frac{1}{\ln |\eta|} \quad {\rm with} \quad
\eta = \frac{1+\cot(\theta^*/2)}{1-\cot(\theta^*/2)},
\label{eq:eta}
\end{equation}
and the coefficient ${\cal C}$ is given by
\begin{equation}
{\cal C} = 2\, \frac{1+|\eta|^\ell}{1-1/\eta^2}.
\end{equation}
Clearly, the result shows that the states $|\Psi^{(\rm even)}_0 \ra$ and $|\Psi^{(\rm odd)}_0\ra$ cannot be distinguished by any local measurement for large $L$. A detailed proof of the inequality Eq. (\ref{eq:expec2}) as well as some explicit examples are given in App. \ref{sec:correlation}. 

\begin{figure}
\centering
\includegraphics[width=0.95\columnwidth]{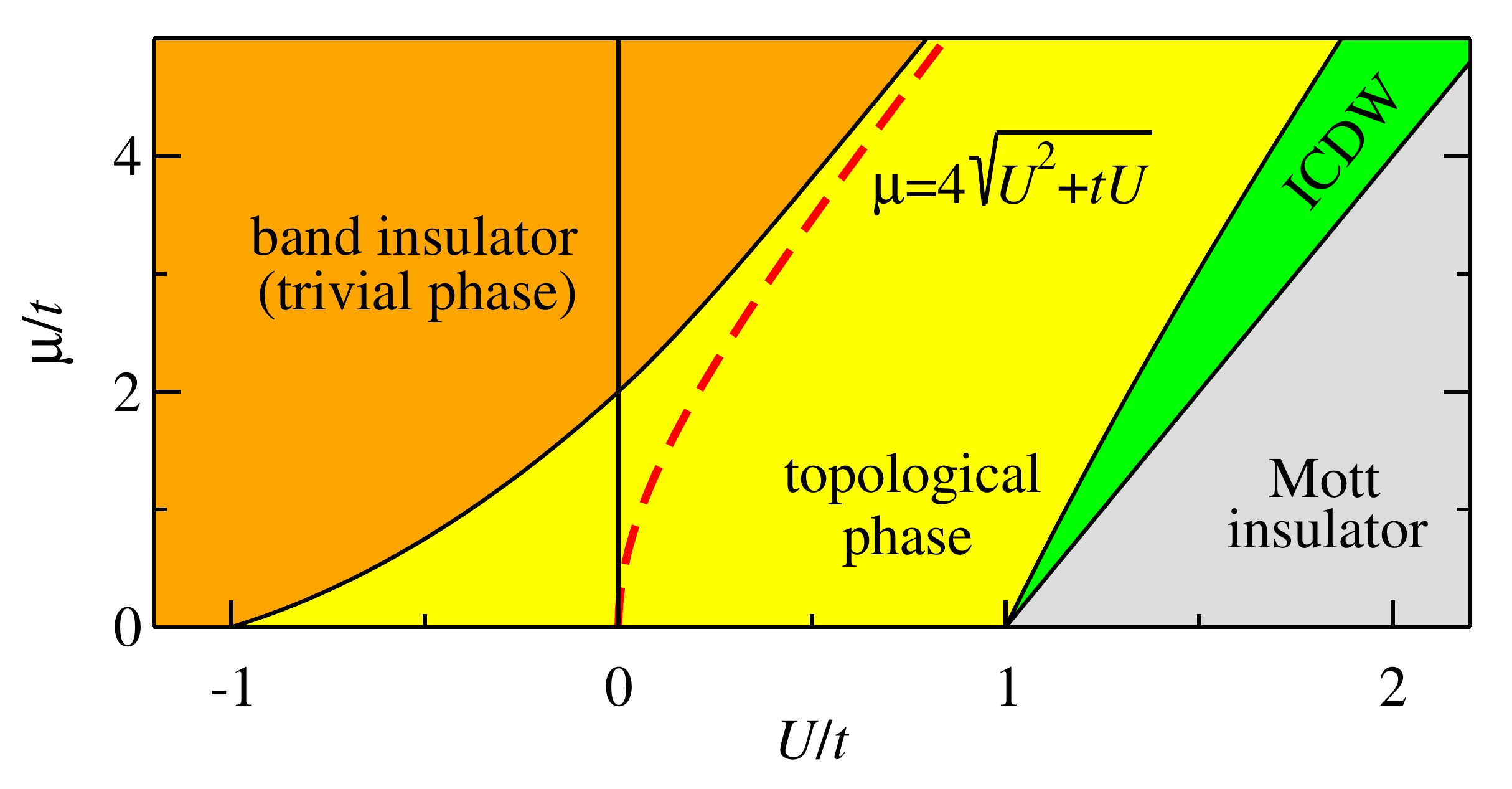}
\caption{(color online). Phase diagram of the interacting Kitaev chain Eq. (\ref{eq:Ham1}) with $\Delta=t$ and $\mu_j=\mu$. The region $\mu/t <0$ is not shown since the phase diagram is invariant under sending $\mu \to -\mu$. The red (dashed) line indicates the frustration-free line Eq. (\ref{eq:FFcond2}), along which the exact ground states can be obtained in closed form. 
ICDW stands for the phase where an incommensurate charge-density wave state is a ground state.}
\label{fig:phase_diagram}
\end{figure}

Let us finally comment on the location of the solvable case Eq. (\ref{eq:FFcond}) in the phase diagram. The phase diagram of the model described by the Hamiltonian Eq. (\ref{eq:Ham1}) with $\Delta=t$ and $\mu_j=\mu$ ($j=1,2,...,L$) has been worked out previously\cite{Sela-11,HS12}; Fig.~\ref{fig:phase_diagram} shows the phase diagram for this particular case. Since the spectrum of the Hamiltonian is invariant under sending $\mu \to -\mu$, we only show the region where $\mu/t \ge 0$. The origin $(U/t, \mu/t)=(0,0)$ corresponds to the original non-interacting Kitaev chain, which is in a topological phase. When $\Delta=t$, the frustration-free condition reads
\begin{equation}
\mu=4 \sqrt{U^2+tU},
\label{eq:FFcond2}
\end{equation}
which is shown by the red (dashed) line in Fig. \ref{eq:Ham1}. 
Clearly, this line is in the topological phase. We note that both the frustration-free line and the phase boundary between the trivial and topological phases approach $\mu=4U$ in the infinite-$U$ limit. 
In this limit, the model can be mapped to a model of hard-core spinless fermions, which is solvable using the free-fermion method.\cite{Gomez_Santos_PRL}

\section{Topological order}
\label{sec:topo order}

In this section we discuss the topological order of the ground states obtained in the previous section. We explicitly construct a one-parameter family of Hamiltonians that interpolates between interacting and non-interacting Kitaev chains, the latter of which exhibits topological order. The spectral gap above the ground state and the existence of Majorana zero modes that are localized around the edges are proven rigorously.

\subsection{Smooth path connecting interacting and non-interacting Hamiltonians}

Let us introduce the Hamiltonian that interpolates continuously between the interacting and non-interacting model. 
As discussed in the previous section, the ground states $|\Psi^{(\pm)}_0 \rangle$ depend only on $\theta^*=\arctan (2\Delta/\mu^*)$ at $\mu=\mu^*$ (frustration-free case). Therefore, they are also ground states of the following one-parameter family of the Hamiltonians
\begin{eqnarray}
H (s) &=& \sum^{L-1}_{j=1} h_j (s) \quad{\rm with}~ s \ge 0, 
\end{eqnarray}
where the local Hamiltonians are given by
\begin{eqnarray}
h_j (s) \!\!&=&\!\! - (c^\dagger_j c_{j+1} + {\rm h.c.}) 
+(1+s)\sin\theta^* (c_j c_{j+1} + {\rm h.c.}) \nonumber \\
\!\!&-&\!\! (1+s)\cos\theta^* (n_j + n_{j+1}-1) \nonumber \\
\!\!&+&\!\! \frac{s}{2} (2n_j-1) (2n_{j+1}-1) +\left( 1+\frac{s}{2} \right).
\label{eq:hamjs}
\end{eqnarray}
Here we have set $t=1$. The last constant was added so that the ground-state energy is equal to zero. Since we have assumed non-negative $\Delta$ and $\mu^*$, $\theta$ is in $[ 0, \pi/2]$. 
When $\mu=\mu^*$, the original Hamiltonian $H$ in Eq. (\ref{eq:ham1}) is the same as $H(2U)$ up to a trivial constant, which can be verified by direct calculation. On the other hand, since the second last term in Eq. (\ref{eq:hamjs}) vanishes when $s=0$, $H(0)$ reduces to the non-interacting Kitaev Hamiltonian with pairing $\Delta/(1+2U)$ and on-site potential $\mu^*/(1+2U)$ which we denote by $H_0$ in the following. Thus we see that the Hamiltonians $H$ and $H_0$ are smoothly connected to each other. Along the entire path $0 \le s \le 2U$, the ground states remain unchanged.\cite{avconf} This suggests that the two Hamiltonians $H$ and $H_0$ are in the same topological phase as long as the energy gap does not close along the path.

Now we show the existence of an energy gap of $H(s)$. 
To this end, we prove the following

\medskip 

\noindent
\textbf{Theorem:} 
{\it For all $s \ge 0$ and $\theta^* \ne 0$, 
(i) the ground state of the Hamiltonian $H(s)$ is unique up to double degeneracy, 
(ii) $H(s)$ has a uniform (independent of the length of the chain) spectral gap above the ground state. 
}

\medskip

\noindent
The proof of the theorem relies on the following 

\medskip

\noindent
\textbf{Lemma:} 
{\it For $\theta^* \ne 0$, 
(i) the ground state of the non-interacting Hamiltonian $H(0)$ is unique up to double degeneracy, 
(ii) $H(0)$ has a uniform (independent of the length of the chain) spectral gap above the ground state. 
}

\medskip
\noindent
This lemma can be proven by directly computing the spectrum of $H(0)$, 
which will be given in the next subsection. 
We shall now show how the theorem follows from the lemma. 

\medskip

\noindent
{\it Proof of Theorem:}~ 
It is convenient to introduce the following operator
\begin{align}
Q_j & = \frac{1}{2}\cos\frac{\theta^*}{2}\,
(c_j+c_{j+1})(c^\dagger_j-c^\dagger_{j+1})(c^\dagger_j+c^\dagger_{j+1}) 
\label{eq:Qop}
\nonumber \\
&+ \frac{1}{2}\sin\frac{\theta^*}{2}\, (c^\dagger_j-c^\dagger_{j+1})
 (c_j+c_{j+1}) (c_j-c_{j+1}),
\end{align}
which is a sum of triple products of fermion operators.\cite{Qsquare} 
In terms of $Q_j$, $h_j (s)$ in Eq. (\ref{eq:hamjs}) is rewritten as
\begin{equation}
h_j (s) = Q_j Q^\dagger_j + (1+s) Q^\dagger_j Q_j,
\label{eq:ineq0}
\end{equation}
which can be verified by a tedious but straightforward calculation.
The above form of $h_j (s)$ is manifestly positive semidefinite for all $s \ge -1$. It is then clear that $h_j (s) \ge h_j (0)$ for $s \ge0$. Here, we write $A \ge B$ to denote that $A-B$ is positive semidefinite.\cite{Tasaki_PTP}  
Since the inequalities for the local Hamiltonians hold for all $j=1,2, ..., L-1$, we arrive at the inequality
\begin{equation}
H (s) \ge H(0), \quad {\rm for}~ s \ge 0.
\label{eq:ineq1}
\end{equation}
The states $|\Psi^{(\pm)}_0 \rangle$ are annihilated by both $Q_j$ and $Q^\dagger_j$ for all $j$, and hence are ground states of $H (s)$. 

We shall next compare the energy eigenvalues of $H(s)$ and $H(0)$. 
Let $E_n (s)$ be the $n$th eigenvalue of $H(s)$. Here the energy eigenvalues are arranged in non-decreasing order: $0 = E_1 (s) = E_2 (s) \le E_3 (s) \le \ldots \le E_n (s) \le \ldots$. Note that $H(s)$ has at least two zero-energy ground states. 
It follows from min-max principle \cite{Horn} that the inequality Eq. (\ref{eq:ineq1}) implies $E_n (s) \ge E_n (0)$ for all $n$. 
Therefore, the spectral gap is a non-decreasing function of $s$, and the eigenstates of $H(s)$ continuously connected to those of $H(0)$ with positive energy never join the zero-energy manifold with increasing $s$. This, together with the lemma, proves the theorem.\hfill$\blacksquare$

\medskip

Several comments are in order. 
Our theorem implies that the interacting Hamiltonian $H(2U)$ and the non-interacting one $H(0)$ are adiabatically connected to each other without gap closing. 
Thanks to the work of Fidkowski and Kitaev\cite{FidkowskiKitaev11}, this suffices to show that $H(2U)$ is topologically non-trivial, because the non-interacting $H(0)$ is 
in a topological phase.
However, some care is needed here. As shown in Refs. [\onlinecite{FidkowskiKitaev10, FidkowskiKitaev11}], the classification of free-fermion topological phases breaks down in one dimensions in the presence of interactions. The non-interacting Kitaev chain with time-reversal symmetry belongs to the BDI symmetry class, different phases of which are characterized by $\mathbb{Z}$ integers. This $\mathbb{Z}$ classification is broken down to $\mathbb{Z}_8$ in the presence of interactions. 
One might think that there is a possibility that a topologically non-trivial phase is adiabatically connected to the trivial phase with no Majorana edge zero modes via interactions. This is, however, not the case here, because the topological index is basically the number of Majorana edge modes localized at one end of the chain. 
For $H(0)$, the number of Majorana edge zero modes at each edge is $1$, which is apparently different from $0$ modulo $8$. Therefore, the one-parameter family of $H(s)$ with $s \ge 0$ is in a topologically non-trivial phase which is adiabatically connected to the one in the non-interacting classification. 

We also comment on the stability of the spectral gap against small perturbations. 
Let us first consider the effect of the inhomogeneity arising from the on-site potential term. 
As noted in Sec. \ref{sec:gs}, the on-site potentials at the boundary sites are half the bulk one in the Hamiltonian Eq. (\ref{eq:ham1}). To see whether the spectral gap is robust against the perturbation that restores the homogeneity to some extent, consider the perturbed Hamiltonian $H' = H +V$ with $\mu=\mu^*$ and
\begin{equation}
V = -\frac{\delta\mu}{2} (n_1 + n_L), 
\end{equation}
where $\delta \mu$ is assumed to be positive. 
Let $E_n (2U, \delta \mu)$ be the $n$th eigenvalue of $H'$. The condition for the existence of the spectral gap reads $E_3 (2U, \delta \mu) > E_2 (2U,\delta \mu)$. From Weyl's theorem (see Theorem 4.3.1 in Ref. \onlinecite{Horn}), one finds $E_n (2U) - \delta \mu \le E_n (2U, \delta\mu) \le E_n (2U)$ for all $n$. This implies that $H'$ has a spectral gap if $\delta \mu < E_3 (2U)$. Since our Theorem ensures that $E_3 (2U)$ is strictly positive, we conclude that the spectral gap is robust against $V$ for small enough $\delta \mu$. A more quantitative estimate is possible, because the lower bound for $E_3(2U)$ is obtained as $E_3 (2U) \ge E_3 (0) = 2 (1-\cos \theta^*)$ in the next subsection. Using this value, we find that the spectral gap of $H'$ is non-zero provided
\begin{equation}
\delta \mu < 2 \left( 1- \frac{\sqrt{(1+2U)^2-\Delta^2}}{1+2U} \right).
\end{equation}
A more interesting question is whether the spectral gap of $H(s)$ is robust against small but global perturbations. For quantum spin systems, rigorous perturbation theories for the stability of gapped ground states have been developed.\cite{Kennedy_Tasaki, Yarotsky} 
Though the existing methods usually assume translation invariant Hamiltonians and do not immediately apply to the present system, we expect that some modification of them is likely to apply particularly to the region around the point $(\Delta, U)=(0,0)$ corresponding to the classical Ising chain. A thorough analysis is, however, beyond the scope of the present study and is left for future work. 

Finally, we comment on the generalization of our model to the case where the couplings are allowed to vary spatially. 
One can, in fact, construct such a model without changing the ground states of the Hamiltonian $H(s)$. 
The key to the construction of such a model is to note that a sum of operators each of which is a polynomial in $Q_j$ and $Q^\dagger_j$ ($j=1,2,..., N-1$) annihilates the states $|\Psi^{(\pm)}_0\rangle$. If this sum is positive semi-definite, then these states are the lowest eigenstates of the sum with zero eigenvalue. 
One such example is the following Hamiltonian:
\begin{align}
H^{\rm inh} = \sum^{L-1}_{j=1} 
(\alpha_j  Q_j Q^\dagger_j + \beta_j Q^\dagger_j Q_j),
\label{eq:Hinh}
\end{align}
where $\alpha_j$ and $\beta_j$ are arbitrary positive numbers. 
Each local interaction can be either repulsive or attractive, depending on $\beta_j/\alpha_j$. More precisely, it is repulsive (attractive) if $\beta_j/\alpha_j >1$ ($0 < \beta_j/\alpha_j < 1$). 
For the above Hamiltonian, we can prove a result similar to the theorem above, and hence the presence of topological order. 
A precise statement and its proof are provided in Appendix \ref{sec:inhomo}. We emphasize that the above example demonstrates that the couplings in the Hamiltonian need not necessarily be uniform to obtain topological order.  

\subsection{Proof of Lemma}
\label{sec:lemma}

Let us prove the lemma by explicitly computing the spectrum of $H(0)$. 
Though $H(0)$ is quadratic in $c$ and $c^\dagger$, it is not quite straightforward to diagonalize it because of the boundaries. Remarkably, however, we find that the entire spectrum of $H(0)$ can be obtained analytically. To show this, we first rewrite $H(0)$ in terms of Majorana operators $a_j=c_j+c^\dagger_j$ and $b_j = (c_j-c^\dagger_j)/\ii$, the properties of which are summarized in App.~\ref{sec:Majo Ham}. 
The Hamiltonian reads
\begin{align}
H (0) = \frac{\ii}{2} \sum^L_{j,k=1} B_{j,k} a_j b_k + (L-1),
\end{align}
where the real matrix $B$ is given by
\begin{align}
B= - \begin{pmatrix}
\fc & 1-\fs \\
1+\fs & 2\fc & 1-\fs \\
 & \ddots & \ddots & \ddots \\
& & 1+\fs & 2\fc & 1-\fs \\
& & & 1+ \fs & \fc
\end{pmatrix}.
\label{eq:Bmat}
\end{align}
with $\fs=\sin \theta^*$, $\fc=\cos \theta^*$. 
Here, matrix elements which are zero are left empty. 
Since the matrix $B$ is non-hermitian, it may not be diagonalizable, 
but can be written in the singular-value decomposition (SVD) form $B = U \Lambda V^{\rm T}$ with $\Lambda = {\rm diag} (\epsilon_1, \epsilon_2, ..., \epsilon_L)$, where orthogonal matrices $U$ and $V$ diagonalize $BB^{\rm T}$ and $B^{\rm T}B$, respectively. 
The singular values $\epsilon_n$ ($n=1,2,..., L$) can be obtained as non-negative square roots of the eigenvalues of  $BB^{\rm T}$. 

To see the relation between these singular values and single-particle energies of $H(0)$,
we introduce the following new set of Majorana operators
\begin{equation}
a'_n = \sum^L_{j=1} U_{j,n} a_j, \quad
b'_n = \sum^L_{j=1} V_{j,n} b_j,
\end{equation}
which satisfy $(a')^\dag_j = a'_j$, $(b')^\dag_j = b'_j$, $\{ a'_j, a'_k \} = \{ b'_j, b'_k \} =2 \delta_{j,k}$, and $\{ a'_j, b'_k \}=0$ for all $j,k=1,2, ... , L$. It then follows from the SVD of $B$ that $H(0)$ is rewritten as 
\begin{align}
H(0) &= \sum^L_{n=1} \epsilon_n a'_n b'_n + (L-1) \\
 &= \sum^L_{n=1} \epsilon_n \left( f^\dagger_n f_n -\frac{1}{2} \right)+(L-1),
\end{align}
where $f_n := (a'_n+ \ii b'_n)/2$ are complex fermions satisfying $\{ f_m, f^\dagger_n \} = \delta_{m,n}$. Therefore, the set of singular values of $B$ is exactly the same as the single-particle spectrum of $H(0)$. To get $\epsilon_n$, we need to diagonalize $BB^{\rm T}$. 
Since this matrix is pentadiagonal, a direct diagonalization by analytical means is not feasible. 
In the present case, however, something special happens. In fact, we have
\begin{equation}
BB^{\rm T} = C^2,
\end{equation}
where $C$ is a real symmetric and tridiagonal matrix given by
\begin{equation}
C= \begin{pmatrix}
1-{\mathfrak s} & {\mathfrak c} \\
{\mathfrak c} & 2 & {\mathfrak c} \\
 & \ddots & \ddots & \ddots \\
& & {\mathfrak c} & 2 & {\mathfrak c} \\
& & & {\mathfrak c} & 1+{\mathfrak s}
\end{pmatrix}.
\label{eq:Cmat}
\end{equation}
A similar factorization was found by Truong and Peschel in the study of the Hamiltonian limit of the corner transfer matrix of the same model.\cite{Truong_Peschel} Because of the presence of the boundary terms, the diagonalization of $C$ is still non-trivial, but can be done analytically. The details are given in App.~\ref{sec:evC}. The exact eigenvalues of $C$ are as follows
\begin{align}
\epsilon_n = \left\{
\begin{array}{ccl}
\displaystyle{2 + 2\fc \cos \left( \frac{n \pi}{L} \right)} & & n=1,2,...,L-1, \\
0 & & n=L.
\end{array}
\right. 
\label{eq:evC}
\end{align}  
The presence of the zero-energy mode clearly signals the topological order in $H(0)$. The system indeed possesses Majorana edge zero modes as we will see in the next subsection. 

Let us now prove the uniqueness of the ground state and the existence of the spectral gap, which does not vanish as $L \to \infty$. From the fact that ${\rm tr}\, C=2 (L-1)$, one finds that $\sum^L_{n=1}\epsilon_n = 2(L-1)$, which together with $\epsilon_L=0$ yields $H(0) = \sum^{L-1}_{n=1} \epsilon_n f^\dagger_n f_n$. 
Since $H(0)$ is positive semi-definite, a state annihilated by $f_n$ for all $n=1,2, ... ,L-1$ is a many-body ground state with zero energy. The states $|\Psi^{({\rm even})}_0 \rangle$ and $|\Psi^{({\rm odd}}_0\rangle$ in Eqs. (\ref{eq:egs}, \ref{eq:ogs}) are such states, because we have already shown that they are zero-energy states of $H(0)$. Then by acting with $f^\dagger_n$ ($n=1,2,...,L-1$) on these two states, one can construct many-body eigenstates of $H(0)$ as follows:
\begin{align}
& (f^\dagger_1)^{m_1} (f^\dagger_2)^{m_2} \cdots (f^\dagger_{L-1})^{m_{L-1}} |\Psi^{({\rm even})}_0\rangle, \\
& (f^\dagger_1)^{m_1} (f^\dagger_2)^{m_2} \cdots (f^\dagger_{L-1})^{m_{L-1}} |\Psi^{({\rm odd})}_0\rangle,
\end{align}
where $m_j=0$ or $1$ for all $j=1,2,...,L-1$. Note that the energy of the above two states is $E = \sum^{L-1}_{j=1} m_j \epsilon_j$, which implies that every energy level is at least two-fold degenerate. 
In this way, we have obtained $2^L$ number of energy eigenstates forming a complete set of orthogonal states, which can be verified by noting $\langle \Psi^{({\rm even})}_0 | \Psi^{({\rm odd})}_0\rangle=0$. 
Because of the positivity of $\epsilon_n$ for $n=1,2,...,L-1$, we find that  $|\Psi^{({\rm even})}_0 \rangle$ and $|\Psi^{({\rm odd}}_0\rangle$ are the only zero-energy ground states of $H(0)$. This proves the uniqueness part of the lemma. 
We now turn to prove the existence of the spectral gap. The lowest-energy excited states are obtained by filling the lowest positive $\epsilon_n$. More explicitly, they are given by 
$f^\dagger_{L-1}|\Psi^{({\rm even})}_0 \rangle$ and $f^\dagger_{L-1}|\Psi^{({\rm odd})}_0 \rangle$. We thus obtain the spectral gap as 
\begin{equation}
E_3 (0) =\epsilon_{L-1}= 
2 \left[ 1-\fc \cos \left( \frac{\pi}{L} \right) \right],
\label{eq:E3}
\end{equation}
which is a decreasing function in $L$ when $L \ge 0$. In the infinite-$L$ limit, we have $E_3 (0) \to 2 (1-\cos\theta^*)$ which is strictly positive unless $\theta^*=0$. This completes the proof of the lemma. \hfill$\blacksquare$

There is an alternative proof of the uniqueness of the two-fold degenerate ground states using the spin-chain Hamiltonian (see Eq. (\ref{eq:spinham}) in App.~\ref{sec:spin Ham}). From Eq. (\ref{eq:spinpara}), one finds that all the off-diagonal elements of the Hamiltonian in the standard Ising basis are non-positive when $\Delta \ge 0$. Then, noting that the space of eigenstates is separated into two disconnected sectors with opposite parities ($\prod^L_{j=1}\sigma^z_j$), we find that in each sector the Hamiltonian satisfies the connectivity condition when $\Delta > 0$, i.e., $\theta^* \ne 0, \pi$. It then follows from the Perron-Frobenius theorem \cite{Horn} that the ground state of each sector is unique. Therefore, the absolute ground state is at most two-fold degenerate. But since we have already found two ground states, this proves that the ground-state degeneracy is exactly two. 

\subsection{Majorana edge zero modes}
\label{sec:zero modes}

The presence of a zero-energy single-particle state is usually a signal of topological order, but does not necessarily mean the existence of Majorana zero modes that are localized at the boundaries. Here we show that, for $H(0)$, the explicit expressions for the Majorana operators can be obtained analytically. Let us first define {\it Majorana edge zero modes}.\cite{Fendley12} A Majorana edge zero mode $\Gamma$ is an operator such that
\begin{itemize}
\item $\Gamma^\dagger = \Gamma$
\item $[H(0), \Gamma] =0$
\item $\{ (-1)^F, \Gamma \}=0$
\item localized near the edge and normalizable as $\Gamma^2 =1$ even in the infinite-$L$ limit.
\end{itemize}
We next show that one can construct such edge modes from the left and right null vectors of $B$ in Eq. (\ref{eq:Bmat}). To see this let ${\bm u} = ( u_1, u_2, ..., u_L )$ be a left null vector of $B$. An elementary calculation shows that 
\begin{equation}
{\bm u} = (1, r, r^2, \cdots, r^{L-1}) 
\quad {\rm with}~ r= - \frac{\fc}{1+\fs},
\end{equation}
satisfies ${\bm u} B = (0,0, \cdots, 0)$. Then from the anticommutation relations of $a$ and $b$ operators, one finds that the following operator commutes with $H(0)$ 
\begin{equation}
\Gamma_{\rm L} = {\cal N} \sum^L_{j=1} r^{j-1} a_j,
\end{equation}
where the normalization factor ${\cal N}$ has been introduced so that $(\Gamma_{\rm L})^2=1$. Similarly, from the right null vector of $B$, one finds that
\begin{equation}
\Gamma_{\rm R} = {\cal N} \sum^L_{j=1} r^{L-j} b_j
\end{equation}
commutes with $H(0)$. Since both $\Gamma_{\rm L}$ and $\Gamma_{\rm R}$ are linear in $a$ and $b$ operators, they are manifestly hermitian and anticommute with the fermionic parity $(-1)^F$. They are localized near the edges and normalizable, because their amplitudes decay exponentially with distance from the boundary unless $\theta^* = 0$. Therefore, the operators $\Gamma_{\rm L}$ and $\Gamma_{\rm R}$ satisfy all the criteria, and hence are Majorana edge zero modes. A linear combination of $\Gamma_{\rm L}$ and $\Gamma_{\rm R}$ gives a non-local complex fermion which is nothing but $f_L = (\Gamma_{\rm L}+\ii \Gamma_{\rm R})/2$, corresponding to the zero-energy state of the single-particle Hamiltonian. Then from the properties of Majorana edge zero modes, together with the uniqueness of the ground states of $H(0)$, it follows that either $\Gamma_{\rm L}$ or $\Gamma_{\rm R}$ maps one ground state to the other with different fermionic parity. 

A special feature of $H(0)$ that is non-interacting {\it and} frustration-free is that the obtained Majorana edge zero modes exactly commute with $H(0)$ even in finite-size systems. This nice property no longer holds for $H(s)$ with $s >0$. Nevertheless, they still map one of the ground states of $H(s)$ to the other since the ground states remain unchanged with the introduction of $s$. Thus, they can be regarded as an {\it interacting generalization of Majorana edge zero modes}. This is one of the main results of our paper. 
We note that one can in principle construct an operator that exactly commutes with $H(s)$ from the adiabatic continuation of $\Gamma_{\rm L}$ or $\Gamma_{\rm R}$.\cite{Hastings_Wen05, Aris_qa} Though this requires a full diagonalization of $H(s)$ and is not feasible, we expect that the resulting operator which is no longer linear in $a$ and $b$ has a significant overlap with $\Gamma_{\rm L}$ or $\Gamma_{\rm R}$. We leave a quantitative analysis for future work.

\section{Concluding remarks}
\label{sec:conc}
In this paper, we have studied the one-dimensional Kitaev chain with nearest-neighbor repulsive interactions. We have shown that the Hamiltonian of the model is frustration-free when the on-site (chemical) potential is tuned to a particular function of the other parameters. Under this condition, the exact ground states can be obtained in closed form. We have also introduced a smooth path between the interacting and non-interacting Kitaev chains, along which the ground states remain unchanged. We proved the following theorem about this one-parameter family of Hamiltonians along this path: (i) the ground state is unique up to double degeneracy, and (ii) there exists a uniform (independent of the chain length) spectral gap above the ground state. 
The theorem implies that the interacting Kitaev chain in the frustration-free case is topologically non-trivial, because it is adiabatically connected to the non-interacting Kitaev chain in the topological phase. 
The stability of the spectral gap against perturbations at the boundary sites was also discussed. Furthermore, we have demonstrated explicitly that there exists a set of Majorana edge zero modes each of which maps one of the ground states to the other with the opposite fermionic parity. 
It would be interesting to explore the implications of this interacting generalization of Majorana edge zero modes in transport properties such as the tunneling conductance. 
Though the dynamical Green functions cannot be obtained analytically even for the frustration-free case, one might be able to develop an efficient computational method using the information of the exact ground states. 
It would also be interesting to see how the results obtained translate into the language of continuum field theories. As is well known, an appropriate scaling limit of the XYZ spin chain is described by the sine-Gordon quantum field theory.\cite{Luther} We thus expect that the continuum limit of the interacting Kitaev chain is also described by a sine-Gordon type model with boundaries. We also expect that boundary bound states\cite{Ghoshal_Zam} arising in integrable quantum field theories are natural candidates for the interacting generalization of Majorana edge zero modes in the continuum limit. The precise correspondence is, however, beyond the scope of the present study and is left for future work.

\section*{Acknowledgment}
The authors thank Alexander V. Balatsky, Lars Fritz, Niklas Gergs, Fabian Hassler, 
Tohru Koma, and Takahiro Morimoto for valuable discussions. 
We especially thank Akinori Tanaka for sharing his results on a related model of interacting spinful fermions \cite{A_Tanaka15} before publication.  
This work was supported in part by JSPS Grants-in-Aid for Scientific Research No. 23740298, 25400407, and 25103007. D.S. acknowledges support of the D-ITP consortium, a program of the Netherlands Organisation for Scientific Research (NWO) that is funded by the Dutch Ministry of Education, Culture, and Science (OCW).

\appendix
\section{Majorana Hamiltonian}
\label{sec:Majo Ham}

For each site $j$, we define the Majorana fermions by
\begin{equation}
a_j := c_j + c^\dag_j,~~~b_j := (c_j-c^\dag_j)/\ii.
\end{equation}
One can easily see that they satisfy the defining relations: 
\begin{align}
a^\dag_j = a_j, \quad b^\dag_j &= b_j, 
\quad \{ a_j, b_k \} =0, \nonumber \\
\{ a_j, a_k \} &= \{ b_j, b_k \} =2 \delta_{j,k}, \nonumber
\end{align}
for all $j,k=1,2, ... , L$. 
When written in terms of $a_j$ and $b_j$, the Hamiltonian $H$ in Eq. (\ref{eq:Ham1}) becomes
\begin{align}
H &= \frac{\ii}{2} \sum^{L-1}_{j=1}
      \left[ (t+\Delta) b_j a_{j+1} - (t-\Delta) a_j b_{j+1} \right] \nonumber\\
   &- \frac{\ii}{2} \sum^L_{j=1} \mu_j a_j b_j 
      -U \sum^{L-1}_{j=1} a_j b_j a_{j+1} b_{j+1} +{\rm const.}
      \label{eq:Majo Ham}
\end{align}
Clearly, the last term represents a quartic interaction 

\section{Spin-chain Hamiltonian and the ground states}
\label{sec:spin Ham}

In this appendix we discuss the relation between the Hamiltonian $H$ in Eq. (\ref{eq:Ham1}) and the spin-chain Hamiltonian. From the standard Jordan--Wigner transformation,\cite{Gangadharaiah-11} 
\begin{equation}
c_1 = \frac{\sigma^x_1 + \ii \sigma^y_1}{2}, \quad
c_j = \left( \prod^{j-1}_{k=1} \sigma^z_k \right)
\frac{\sigma^x_j + \ii \sigma^y_j}{2}~~{\rm for}~2 \le j \le L,
\nonumber
\end{equation}
we obtain explicit expressions of Majorana fermions in terms of spin operators as
\begin{align}
a_1 &= \sigma^x_1,~~a_j = \left( \prod^{j-1}_{k=1} \sigma^z_k \right) \sigma^x_j~~{\rm for}~2 \le j \le L, 
\label{eq:asp} \\
b_1 &= \sigma^y_1,~~b_j = \left( \prod^{j-1}_{k=1} \sigma^z_k \right) \sigma^y_j~~{\rm for}~2 \le j \le L,
\label{eq:bsp}
\end{align}
where $(\sigma^x_j, \sigma^y_j, \sigma^z_j)$ are the Pauli matrices at site $j$. 
We note here that the fermionic parity in spin variables is written as
\begin{equation}
(-1)^F = \prod^L_{j=1} \sigma^z_j.
\end{equation}
Substituting Eqs. (\ref{eq:asp}, \ref{eq:bsp}) into Eq. (\ref{eq:Majo Ham}), we have 
\begin{align}
H &= \sum^{L-1}_{j=1} (-J_x \sigma^x_j \sigma^x_{j+1} -J_y \sigma^y_j \sigma^y_{j+1} 
+ J_z \sigma^z_j \sigma^z_{j+1}) \nonumber \\
&-\sum^L_{j=1} B_j \sigma^z_j +{\rm const.}
\label{eq:spinham}
\end{align}
where the parameters are given by 
\begin{align}
J_x = \frac{t+\Delta}{2}, \; J_y = \frac{t-\Delta}{2}, \;
J_z = U, \; B_j = -\frac{\mu_j}{2}.
\label{eq:spinpara}
\end{align}
Recent theoretical proposals to realize spin-$1/2$ XYZ chain in a magnetic field using optical lattice systems can be found in Refs.~\onlinecite{Pinheiro_PRL, Piraud_PRA, Peotta_JStat} 

In the spin-chain language, the frustration-free condition Eq. (\ref{eq:FFcond}) reads
$B_1 = B_L = B^*/2$ and $B_j =B^*$ ($j=2,3,...,L-1$), where 
\begin{equation}
B^* = -2 \sqrt{(J_z+J_x) (J_z+J_y)}.
\end{equation}
This is exactly what has been found in Refs. \onlinecite{Kurmann-82} and \onlinecite{MuellerShrock85}. 
When written in terms of spin states, the ground states Eq. (\ref{eq:gs1}) become simple direct product states:
\begin{equation}
|\Psi^{(\pm)}_0\rangle = \bigotimes^L_{j=1} 
(|\uparrow\rangle_j \pm \alpha |\downarrow\rangle_j), 
\end{equation}
which can be distinguished by the expectation value of the local operator $\sigma^x_j$. We note, however, that $\sigma^x_j$ for general $j$ is non-local in the fermions, because it involves a string of fermion operators. 
The ground states with fixed (fermionic) parities, i.e., Eqs. (\ref{eq:egs}, \ref{eq:ogs}), are no longer direct products, but can still be written in the form of matrix product states.  Their explicit expressions can be found in Ref.~\onlinecite{Asoudeh_PRL}.

\section{Expectation values of $\boldsymbol{O_\text{e}}$}
\label{sec:correlation}

In this appendix we present a derivation of the inequality Eq. (\ref{eq:expec2}). We also provide explicit expressions for several kinds of correlation functions in the infinite-$L$ limit. 

\subsection{Derivation of Eq. (\ref{eq:expec2})}

Let us first prove the inequality Eq. (\ref{eq:expec2}). Noting that $[\Oe, (-1)^F]=0$ and using the relation $(-1)^F | \Pspm\ra = |\Psmp \ra$, we find
\begin{align}
\la \Psp |\Oe |\Psp \ra = \la \Psm |\Oe |\Psm \ra, \\
\la \Psp |\Oe |\Psm \ra = \la \Psm |\Oe |\Psp \ra.
\end{align}
This yields
\begin{align}
\la \Oe \ra_{\rm even} = 
\frac{\la \Psp|\Oe |\Psp\ra + \la \Psp|\Oe |\Psm\ra}
{1+1/\eta^L}, \\
\la \Oe \ra_{\rm odd} = 
\frac{\la \Psp|\Oe |\Psp\ra - \la \Psp|\Oe |\Psm\ra}
{1-1/\eta^L},
\end{align}
where we have used the fact that
$\la \Pspm|\Pspm \ra =1$ and 
$\la \Pspm|\Psmp \ra  =1/\eta^L$ (see Eq. (\ref{eq:eta}) for the definition of $\eta$). Then we find 
\begin{align}
&\!\!\!\!\!\!\!\!\!\big| \la \Oe \ra_{\rm even} - \la \Oe \ra_{\rm odd}\big|
\nonumber \\
\le &\, 2\, \frac{(1/|\eta|^L)\, \|\Oe \| + \left| \la \Psp |\Oe |\Psm\ra \right|}{1-1/\eta^{2L}},
\label{eq:preineq}
\end{align}
where the following inequality has been used
\begin{align}
\big| \la \Psp| \Oe |\Psp \ra \big| \le \|\Oe \|.
\end{align}
In the following, we focus on $0 < \theta^* < \pi$, which implies $|\eta| >1$. 
To evaluate $\left| \la \Psp |\Oe |\Psm\ra \right|$, we recall that $\Oe$ is a local operator supported on a finite number of sites. Since $c_j$ and $c^\dag_j$ commute with $\Oe$ unless $j=j_1, ..., j_k$, one finds
\begin{align}
\la \Psp |\Oe |\Psm\ra = \frac{1}{\eta^{L-\ell}}\,
\la {\widetilde \Psi}^{(+)}_0 | \Oe| {\widetilde \Psi}^{(-)}_0 \ra,
\end{align}
where
\begin{align}
| {\widetilde \Psi}^{(\pm)}_0 \ra :=
\frac{1}{(1+\alpha^2)^{\ell/2}}\, 
\left( \prod^{j_k}_{j=j_1}
e^{\scalebox{0.95}{$\pm \alpha c^\dag_j$}} \right)
|{\rm vac}\ra,
\end{align}
are the ground states restricted to the $\ell$ consecutive lattice sites $j=j_1, j_1 + 1, ..., j_k$. Then using the Schwartz inequality, one finds 
\begin{align}
\left| \la {\widetilde \Psi}^{(+)}_0 | \Oe| {\widetilde \Psi}^{(-)}_0 \ra \right|^2 
& \le  \la {\widetilde \Psi}^{(+)}_0|  {\widetilde \Psi}^{(+)}_0 \ra \, \la {\widetilde \Psi}^{(-)}_0 | \Oe^\dag \Oe | {\widetilde \Psi}^{(-)}_0 \ra \nonumber \\
& \le \| \Oe \|^2.
\end{align}
Substituting this into Eq. (\ref{eq:preineq}), one finds
\begin{align}
&\!\!\!\!\!\!\!\!\!\big| \la \Oe \ra_{\rm even} - \la \Oe \ra_{\rm odd}\big| 
\le  2\, \frac{1/|\eta|^L + 1/|\eta|^{L-\ell}}{1-1/\eta^{2L}}\, \|\Oe \|
\nonumber \\
\le &\, \frac{2}{|\eta|^L}\times \frac{1+|\eta|^\ell}{1-1/\eta^2}\, \|\Oe \|,
\label{eq:finineq}
\end{align}
where the second line follows from 
\begin{align}
\frac{1}{1-x^L} \le \frac{1}{1-x} \quad
{\rm for} \quad 0 \le x <1.
\end{align}
Then it is obvious that the desired inequality Eq. (\ref{eq:expec2}) follows from Eq. (\ref{eq:finineq}). 

\subsection{Correlation functions}

As we have shown above, any local $\Oe$ has the same expectation value in $|\Pse\ra$ and $|\Pso\ra$ in the infinite-$L$ limit; for completeness we finally give explicit expressions for some of correlation functions. For density operators $n_j$, we have
\begin{align}
& \lim_{L \to \infty} \langle n_j \rangle_{\rm even/odd} 
= \frac{1}{1+\tan (\theta^*/2)},
\\
& \lim_{L\to \infty} \langle n_j n_k \rangle_{\rm even/odd} 
= \left[ \frac{1}{1+\tan (\theta^*/2)} \right]^2.
\end{align}
Similarly one can compute equal-time Green functions. For $j < k$ with $|j-k| < \infty$, we have
\begin{align}
& \lim_{L \to \infty} \langle c_j c^\dagger_k \rangle_{\rm even/odd} \nonumber \\
&=-\frac{\sin\theta^*}{2 (1+\sin \theta^*)}\,
\left[ \frac{1-\cot (\theta^*/2)}{1+\cot (\theta^*/2)} \right]^{k-j-1},
\end{align}
which decays exponentially in the distance $|j-k|$. 

\section{Inhomogeneous model}
\label{sec:inhomo}

In this appendix we show that the model described by the Hamiltonian $H^{\rm inh}$ (Eq. (\ref{eq:Hinh})) exhibits topological order. To this end, we prove the following

\medskip 

\noindent
\textbf{Proposition:} 
{\it For all $\alpha_j >0$, $\beta_j >0$ ($j=1,2,...,L-1$) and $\theta^* \ne 0$, (i) the ground state of the Hamiltonian $H^{\rm inh}$ is unique up to double degeneracy, (ii) $H^{\rm inh}$ has a uniform (independent of the length of the chain) spectral gap above the ground state. 
}

\medskip

\noindent
This can be proven along the same lines as Theorem in the main text. 

\medskip

\noindent
{\it Proof of Proposition:}~
We first introduce the minimum value of the inhomogeneous parameters as
\begin{align}
\gamma = {\rm min}\, \{ \alpha_j \}^{L-1}_{j=1} \cup 
\{ \beta_j \}^{L-1}_{j=1},
\end{align}
which is, by definition, strictly positive ($\gamma > 0$). 
Then from the fact that $Q_j Q^\dagger_j \ge 0$ and $Q^\dagger_j Q_j \ge 0$, we find
\begin{align}
H^{\rm inh} \ge \gamma \sum^{L-1}_{j=1} 
(Q_j Q^\dagger_j + Q^\dagger_j Q_j) 
= \gamma H(0),
\label{eq:inhcomp}
\end{align}
where $H(0)$ is the Hamiltonian of the non-interacting Kitaev chain. The Hamiltonians $H(0)$ and $H^{\rm inh}$ share the same ground states $|\Psi^{(\pm)}_0 \rangle$, because they are annihilated by both $Q_j$ and $Q^\dagger_j$ for all $j$. 

Let us next compare the energy eigenvalues of $H^{\rm inh}$ and $H(0)$. Let $E^{\rm inh}_n$ be the $n$th eigenvalue of $H^{\rm inh}$. Here the energy eigenvalues are arranged in non-decreasing order: $0=E^{\rm inh}_1 = E^{\rm inh}_2 \le E^{\rm inh}_3 \le \ldots \le E^{\rm inh}_n \le \ldots $. It then follows from min-max principle\cite{Horn} that the inequality Eq. (\ref{eq:inhcomp}) implies $E^{\rm inh}_n \ge \gamma E_n (0)$ for all $n$. This immediately implies that $E^{\rm inh}_3 > 0$ since we have already shown that $E_3 (0)$ is strictly positive unless $\theta^* \ne 0$ (see Eq. (\ref{eq:E3})). This completes the proof of property (ii). 
Property (i) also follows from min-max principle and Lemma in Sec. IV A.  \hfill$\blacksquare$

\medskip

Let us next show the existence of topological order in the system described by $H^{\rm inh}$. We can construct the Hamiltonian that interpolates between $H^{\rm inh}$ and $H(0)$ as follows:
\begin{align}
H^{\rm inh} (x) = x H^{\rm inh} + (1-x) \gamma H(0),    
\end{align}
where $0 \le x \le 1$. Using the inequality Eq. (\ref{eq:inhcomp}), one can check that $H^{\rm inh} (x) \ge \gamma H(0)$ for $0 \le x \le 1$. Therefore, by repeating the proof of Proposition, one finds that the spectral gap of $H^{\rm inh}(x)$ does not close along the path connecting $\gamma H(0)$ and $H^{\rm inh}$. This, together with the result of Fidkowski and Kitaev\cite{FidkowskiKitaev11}, implies the existence of topological order in the system. 

Let us remark that a local interaction between the neighboring sites can be either repulsive or attractive in $H^{\rm inh}$, depending on $\beta_j/\alpha_j$. To see this, we rewrite the Hamiltonian as
\begin{align}
H^{\rm inh} = \sum^{L-1}_{j=1} \alpha_j\, 
h_j \left( \frac{\beta_j}{\alpha_j}-1 \right),
\end{align}
where $h_j (s)$ is given by Eq. (\ref{eq:hamjs}). 
It is then clear that the local interaction between $n_j$ and $n_{j+1}$ is repulsive if $\beta_j/\alpha_j > 1$, while it is attractive if $0 < \beta_j/\alpha_j < 1$. The interactions can be made purely attractive by taking a set of $\alpha_j$ and $\beta_j$ that satisfy $0 < \beta_j/\alpha_j < 1$ for all $j$. For instance, if the parameters are chosen as
\begin{align}
\alpha_j = 1, ~~~ \beta_j = 1+2U ~~~ (j=1,2, ..., L-1),
\end{align}
the Hamiltonian $H^{\rm inh}$ reduces to $H(2U)$, which describes the Kitaev chain with attractive interactions when $U<0$. Note that the condition $U > -1/2$ is required so that $H(2U)$ is positive semidefinite. 

\section{Eigenvalues of the matrix $C$}
\label{sec:evC}

In this appendix we present a detailed exposition of the calculation of the eigenvalues of $C$ in Eq. (\ref{eq:Cmat}). We first make an ansatz for the eigenvectors. Let ${\bm v}(q) = (v_1 (q), v_2 (q), ..., v_L (q))$ be 
an eigenvector of $C$. For each component, we choose the following ansatz
\begin{equation}
v_j (q) = \alpha e^{\ii q j} + \beta e^{-\ii q j},
\label{eq:ansatz}
\end{equation}
where the coefficients $\alpha$, $\beta$, and the ``wavenumber'' $q$ will be determined from the matching conditions at the boundaries. 
The eigenvalue equation $C {\bm v}(q) = \epsilon (q) {\bm v}(q)$ can be written in components as 
\begin{align}
& (1-\fs) v_1 (q) +\fc v_2 (q) =\epsilon (q) v_1 (q), 
\label{eq:Sch1}\\
& \fc v_{j-1} (q) + 2 v_{j} (q) + \fc v_{j+1} (q) = v_j (q),\; 1<j<L, 
\label{eq:Sch2}\\
& \fc v_{L-1} (q) + (1+\fs) v_L (q) = \epsilon (q) v_L.
\label{eq:Sch3}
\end{align}
Substituting Eq. (\ref{eq:ansatz}) into Eq. (\ref{eq:Sch2}), one finds that the eigenvalue must be of the form:
\begin{equation}
\epsilon (q) = 2 + 2\fc \cos q.
\label{eq:evq}
\end{equation}
Then, from the matching conditions at the boundaries, i.e., Eqs. (\ref{eq:Sch1}, \ref{eq:Sch3}), the consistency condition for $q$ is obtained as
\begin{align}
\begin{pmatrix}
-\fc -(1+\fs) e^{\ii q}\, & -\fc -(1+\fs)e^{-\ii q} \\
(\fs -1 -\fc e^{\ii q}) e^{\ii qL}\, & (\fs -1 -\fc e^{-\ii q}) e^{-\ii qL}
\end{pmatrix}
\begin{pmatrix}
\alpha \\ \beta
\end{pmatrix}
=
\begin{pmatrix}
0 \\ 0
\end{pmatrix}. \nonumber
\end{align}
The above equation has a non-trivial solution if the determinant of the matrix in the left-hand side vanishes. This condition reads
\begin{equation}
-4 \ii \fc (1+\fc \cos q) \sin (qL) = 0. 
\label{eq:det}
\end{equation}
The real wavenumbers are then obtained as
\begin{equation}
q = \frac{n \pi}{L}, \quad n=1,2, ...., L-1.
\end{equation}
Note that $q=0$ and $\pi$ are not allowed because they result in ${\bm v}(q)={\bm 0}$. The equation (\ref{eq:det}) also has a complex solution 
\begin{equation}
q = q_0 = \pi + \ii \ln \left( \frac{\fc}{1+\fs} \right),
\end{equation}
which is a solution of $\cos q_0 = -1/\fc$. One might think that the set of solutions obtained is overcomplete because $q=-q_0$ is a solution as well. However, this is not the case since both $q_0$ and $-q_0$ yield the same ${\bm v}(q)$. We finally obtain the desired eigenvalues Eq. (\ref{eq:evC}) by substituting the obtained $q$ into Eq. (\ref{eq:evq}).


\begin{thebibliography}{150}
\bibitem{RefB}
This can also be seen by noting that each local Hamiltonian $(+ {\rm const}.)$ can be cast in the form Eq. (\ref{eq:ineq0}) at $\mu=\mu^*$ and $|\Psi^{(\pm)}_0\rangle$ are annihilated by both $Q_j$ and $Q^\dagger_j$ for all $j$. For the definition of $Q_j$, see Eq. (\ref{eq:Qop}).
\bibitem{avconf}
This path has no relation to the exactly solvable line shown in Fig. \ref{fig:phase_diagram}.
\bibitem{Qsquare}
We remark that each of $Q_j$ and $Q^\dagger_j$ squares to zero.
\bibitem{Aris_qa}
This program has been carried out for a slightly different model of interacting Majorana fermions, corresponding to the XYZ spin chain without magnetic field.\cite{Aris-15}
\end{thebibliography}
\end{document}